\pdfoutput=1
\documentclass[final,times,1p]{elsarticle}

\usepackage{graphics}
\usepackage{graphicx,stfloats}
\usepackage{color}

\usepackage{amssymb}

\usepackage{amsthm}
\usepackage{amsmath}
\usepackage{tabularx}
\usepackage{mathtools} 
\usepackage{latexsym}
\usepackage{subfig}
\usepackage{float}
\usepackage{url}
\usepackage{setspace}
\usepackage{multirow}
\usepackage{threeparttable}
\usepackage[version=4]{mhchem}
\usepackage{ulem} 
\biboptions{sort&compress}

\journal{Combustion and Flame}
\newcommand{\ode}[2]{\frac{\text{d} {#1}}{\text{d} {#2}}}
\begin{document}
\begin{frontmatter}
	
\title{Non-equilibrium effects on thermal ignition using hard sphere molecular dynamics}

\author[add1]{R. Murugesan\corref{cor1}}
\ead{rmuru024@uottawa.ca}
\author[add1]{N. Sirmas}
\author[add1]{M. I. Radulescu}

\address[add1]{Department of Mechanical Engineering, University of Ottawa, Ottawa, \\Ontario K1N 6N5, Canada}
\cortext[cor1]{Corresponding author:}

\begin{abstract}
The present study addresses the role of molecular non-equilibrium effects in thermal ignition problems.  We consider a single binary reaction of the form A+B$\rightarrow$C+C.  Molecular dynamics calculations were performed for activation energies ranging between $RT$ and 7.5$RT$ and heat release of 2.5$RT$ and 10$RT$.  The evolution of up to 10,000 particles was calculated as the system undergoes a thermal ignition at constant volume.  Ensemble averages of 100 calculations for each parameter set permitted to determine the ignition delay, along with a measure of the stochasticity of the process.  A well behaved convergence to large system sizes is also demonstrated.  The ignition delay calculations were compared with those obtained at the continuum level using rates derived from kinetic theory: the standard rate assuming that the distribution of the speed of the particles is the Maxwell-Boltzmann distribution, and the perturbed rates by Prigogine and Xhrouet \cite{Prigogine_1949} for an isothermal system, and Prigogine and Mahieu \cite{Prigogine_1950} for an energy releasing reaction, obtained by the Chapman-Enskog perturbation procedure.  The molecular results were found in very good agreement with the latter at high activation energies (low temperature), confirming that non-equilibrium effects promote the formation of energetic particles, that serve as seeds for subsequent reaction events: i.e., molecular hot spots.  This effect was found to lower the ignition delay by up to 30\%.  At low activation energies (high temperatures), the ignition delay obtained from the standard equilibrium rate was found to be up to 60\% longer than the molecular calculations.  This effect is due to the rapidity of the reactive collisions that do not allow the system to equilibrate.  For this regime, none of the perturbation solutions obtained by the Chapman-Enskog procedure were valid.  This study thus clearly shows the importance of non-equilibrium effects in thermal ignition problems, for most activation energies of practical interest. These non-equilibrium effects are discussed in the context of hydrogen auto-ignition at low temperatures.  The auto-catalytic loops of H$_2$O$_2$ and HO$_2$ controlling the thermal ignition process, which have both high energy release production of these species subsequently involved in high activation reactions of these species, can be mostly impacted by hotspot non-equilibrium effects.
\end{abstract}


\end{frontmatter}

\section{Introduction}
\label{Introduction}
Thermal ignition of a reactive gas involves the classical coupling between the temperature of a reactive mixture of gases and the rate of energy addition, which depends exponentially on temperature. A slight increase in the system's temperature by reactions may bring the system in a runaway reaction occurring faster than any potential loss \cite{Williams_1985}. Although real systems involve the evolution of many reactive species in a complex sequence of elementary reactions, the thermal ignition problem is usually addressed with a single reaction with Arrhenius kinetics.  Thermal ignition is particularly relevant to low temperature ignition in hydrogen, for example \cite{Sanchez_2014}.  The reduction of a multi-step chemical system to a one or two step formalism relevant to auto-ignition follows classical arguments \cite{Sanchez_2014}, which we address later in the paper. 

A proper account of the reaction rate of elementary reactions is essential in problems of thermal ignition.  Although most reaction rates of elementary reactions are obtained or calibrated from experiment, their assumed Arrhenius form is justified theoretically from arguments of kinetic theory of gases or transition state theory only if the gas can be assumed to be in thermodynamic equilibrium, i.e., the distribution of molecular speeds follows the Maxwell-Boltzmann distribution.  While the assumption of a Maxwell-Boltzmann distribution is expected asymptotically in the limit of slow reactions, i.e., when the activation energy is large and the heat release is negligible, deviations from the Maxwell-Boltzmann (MB) distribution are expected otherwise. 

The problem of non-equilibrium associated with reactive systems has gained considerable amount of attention in the literature for a long time \cite{Prigogine_1949,Borisov_1974,Gorecki_1987,Gorecki_1991,Gorecki_2000,Lemarchand_2004,Present_1968,Shizgal_1970,Shizgal_1996}. Most of the researchers employed the Chapman-Enskog method of solution of the Boltzmann equation \cite{Chapman_1958}. On the other hand, in the mid 80's Dean and Westmoreland considered the need for including the chemically activated states in association and dissociation reactions to the kinetic models using QRRK (Quantum Rice-Ramsperger-Kassel) theory \cite{Dean_1985, Westmoreland_1986}. They also studied the temperature and pressure dependence of the rate constants of individual reactions due to these chemically activated intermediates. For some of the elementary reactions in combustion, they identified that the radical production reaction in high temperature and the recombination reaction in low temperature plays a crucial role in combustion relevant conditions and concluded that these reactions become more pressure-dependent at high pressures. In this context, in recent years many researchers \cite{Dontgen_2017,Dontgen_Feb2017, Burke_2015, Labbe_2017, Goldsmith_2015} have investigated the non-equilibrium effects specific to low-temperature combustion processes. Burke et al. proposed that the non-Boltzmann reaction sequence involved in low temperature chain branching process significantly affect the evolution of the system \cite{Burke_2015}. Labbe et al. studied the impact of including non-equilibrium effects for the formyl radical, HCO, in simulations of laminar flame speeds in which HCO has been a prescient indicator for heat release in combustion of hydrocarbons and predicted 7-13\% increase in laminar flame speed \cite{Labbe_2017}. In a similar manner, Goldsmith et al. focused on the non-Boltzmann effects of low-temperature chemistry of propane oxidation and concluded that these effects can affect the kinetics of subsequent reactions \cite{Goldsmith_2015}. On the other hand, D$\ddot{\text{o}}$ntgen et al. \cite{Dontgen_2017, Dontgen_Feb2017} investigated the new pathway of $beta$-scission of rotationally excited radicals formed through hydrogen abstraction. They reported that hot $beta$-scission was not only found to increase the rate of dissociation, but also dominates the fuel radicals at temperature above approximately 1050 K and pressures up to 100 atm.

Prigogine and co-workers were the first to consider the non-equilibrium effects in single reactions by extending the classical perturbation method of Chapman and Enskog to the case of reactive collisions of hard spheres \cite{Prigogine_1949, Prigogine_1950}.  They investigated two separate problems: the non-equilibrium effects arising in an isothermal system \cite{Prigogine_1949} and in a reactive system, where the energy release may influence the rate of reactions explicitly by perturbing the spatially local MB distribution \cite{Prigogine_1950}.  The latter is particularly important for the problem of interest in the present paper, namely thermal ignition.     In the isothermal case, the non-equilibrium corrections to the MB-derived rate of standard kinetic theory of gases was found to be small in such Chapman-Enskog perturbations to the MB distribution \cite{Prigogine_1949}. Subsequent work by other groups confirmed this finding \cite{Present_1968, Shizgal_1970}, although problems have been raised regarding the uniformity of the Chapman-Enskog expansion.  The advent of affordable calculations using molecular dynamics (MD) or Direct Simulation Monte Carlo (DSMC) techniques \cite{Gorecki_2000,Gorecki_1987, Lemarchand_2004, Mansour_1992} permitted to evaluate the predictions of the kinetic theory results.  Although a systematic comparison has not been performed over an entire parameter range of activation energies and heats of reaction, the computational results confirmed that the corrections in the isothermal case tend to slow down the reactions.   More interestingly, however, is the reactive case treated in the perturbation theory of Prigogine and Mahieu. Their results indicate that the effect of heat release significantly accelerate the reaction rate through non-equilibrium effects. That is, the reactive collisions through local energy addition give rise to highly energetic particles that serve to initiate other reactive collisions locally, before the effect is equilibrated in the entire system. In this case, the perturbation is proportional to the ratio $Q/E_{\text{A}}$, where $Q$ and $E_{\text{A}}$ are the heat release and activation energy, in that the energetic systems will benefit the most of these effects.  

Recently, more relevant to the present study, Sirmas et al.\ have investigated the role of non-equilibrium effects in the problem of thermal ignition in a model system characterized by a single binary reaction by molecular dynamics in two-dimensions \cite{Sirmas_2017}.  They have identified large departures from the equilibrium prediction at both low and high activation energies. It was observed that systems with low activation energies exhibit a homogeneous ignition event with departure from thermal equilibrium prediction, yielding lower reaction rates and longer ignition delays. On the other hand, for systems with sufficiently high activation energies and heat of reaction, they have proposed that the thermal ignition is a stochastic process with ignition due to hot-spot formations which is associated with shorter ignition delay than predicted from continuum model by approximately 30\%.  While these results pointed to the importance of non-equilibrium effects, the 2D problem can be argued to display a stronger propensity for non-equilibrium than 3D because of the reduction in dimensionality.  In 1D for example, the gas never relaxes to a Maxwell-Boltzmann distribution.

In the present communication, we wish to assess the role of non-equilibrium effects in thermal ignition in a much more physical three-dimensional space, in order to compare with the analytical predictions of non-equilibrium kinetic theory for both isothermal and reactive collisions.  We consider the same single binary reaction as Sirmas et al. in a hard sphere gas.  The assumption of the hard sphere gas permits to not only compare with the available predictions from kinetic theory, but also to integrate the equations of motion of the spheres exactly and deterministically, given a set of initial conditions of positions and momenta. We make use of the classical algorithm of Alder and Wainright to evolve the system analytically, now known as the Event Driven Molecular Dynamics method \cite{Alder_1959}. It is to be noted that for a dilute gas, relevant to ideal gases, that a hard sphere potential yields accurate results \cite{Vincenti_1975}, and is therefore chosen due to its adaptability into an event driven algorithm which significantly lowers the computational time. These simplifying assumptions must be re-evaluated when treating a real system. However, the simplifications used unambiguously identify various mechanisms for departures from the predictions using the macroscopic models based on local equilibrium.

\section{Problem definition}
\label{Model description}
The problem we solve is the evolution of a system of hard spheres evolving inside an insulated cube of side $L$, i.e., a constant volume ignition problem.  
By hard spheres, we refer to particles which do not exert any force on others except at the instant of collision, where the laws of momentum and energy conservation apply to determine the post-collision velocities \cite{Vincenti_1975}. The gas contains three types of species, A, B and C, which may transform according to the model binary irreversible exothermic reaction:
\begin{align}
 \text{A + B} \rightarrow  \text{C + C} + \text{heat}
\end{align} 
Both reactants and the product have identical mass and diameter, $d$.  At time zero, we take the gas to be in equilibrium, such that the speed distribution is given by the Maxwell-Boltzmann distribution. The initial temperature of the system, defined from the mean speed of the particles, uniquely defines the initial condition in the thermodynamic sense.  With these initial conditions, we allow the particle dynamics inside the box to evolve and undergo an overall thermal ignition process.  
  
\begin{figure*}
	\centering
	\includegraphics[trim={0cm 0.1cm 0.3cm 0.1cm},clip,width=.85\columnwidth]{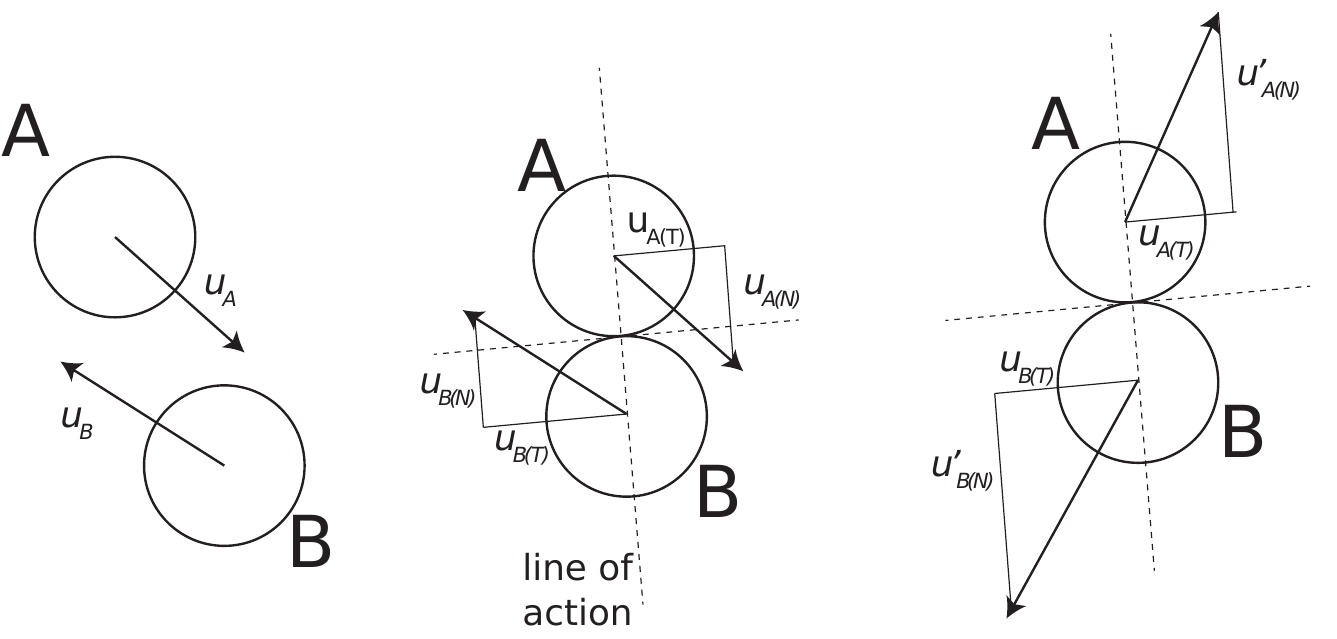}
	\caption{Schematic of interactions between particles A and B before impact (left), during impact (center) and after impact (right) along the line of action \cite{Sirmas_2017}}
	\label{fig:line of centres}
\end{figure*}  
  
All collisions are assumed to be elastic with the exception of reactive collisions. The heat release $Q$ of a reactive collision increases the kinetic energy of each species C.  The collision occurs along the line of action, i.e., the change in the speed can take place only in the normal direction while the tangential components remain unchanged (as seen in Fig. \ref{fig:line of centres}).  The reactive collision can occur only when the relative speed between the two colliding reactants exceeds the minimum impact velocity, $u_{\text{cr}}$, satisfies the condition $\left(|u_{\text{A(N)}}-u_{\text{B(N)}}|> u_{\text{cr}}\right)$.  The impact velocity is related to the activation energy by $u_{\text{cr}}=(4E_\text{A}/m)^{1/2}$.  The post collision speed of particle A (which becomes a particle C) after a reactive collision, is derived from the energy equation of the form:
\begin{align}
\frac{1}{2}m_A u_A^2 + \frac{1}{2} m_B u_B^2 + Q = \frac{1}{2} m_A u_A'^2+ \frac{1}{2} m_B u_B'^2
\label{energy equation}
\end{align}
which can be re-written as
\begin{align}
u_\text{A(N)}'= \frac{1}{2}\bigg(u_{\text{A+B}} + u_{\text{A-B}}\sqrt{1 + \frac{4Q}{m u_{\text{A-B}}^2}}\bigg)
\label{post_collision_velocities}
\end{align}
where, $u_{\text{A-B}}=u_\text{A(N)}-u_\text{B(N)}$ and $u_{\text{A+B}}=u_\text{A(N)}+u_\text{B(N)}$; refer to Fig.\ 1 for notation. The tangential component of post collision velocities remains unchanged as per the line of action model.

\begin{figure*}
	\centering
	\includegraphics[trim={0.1cm 0.1cm 0.1cm 0.25cm},clip,width=.7\columnwidth]{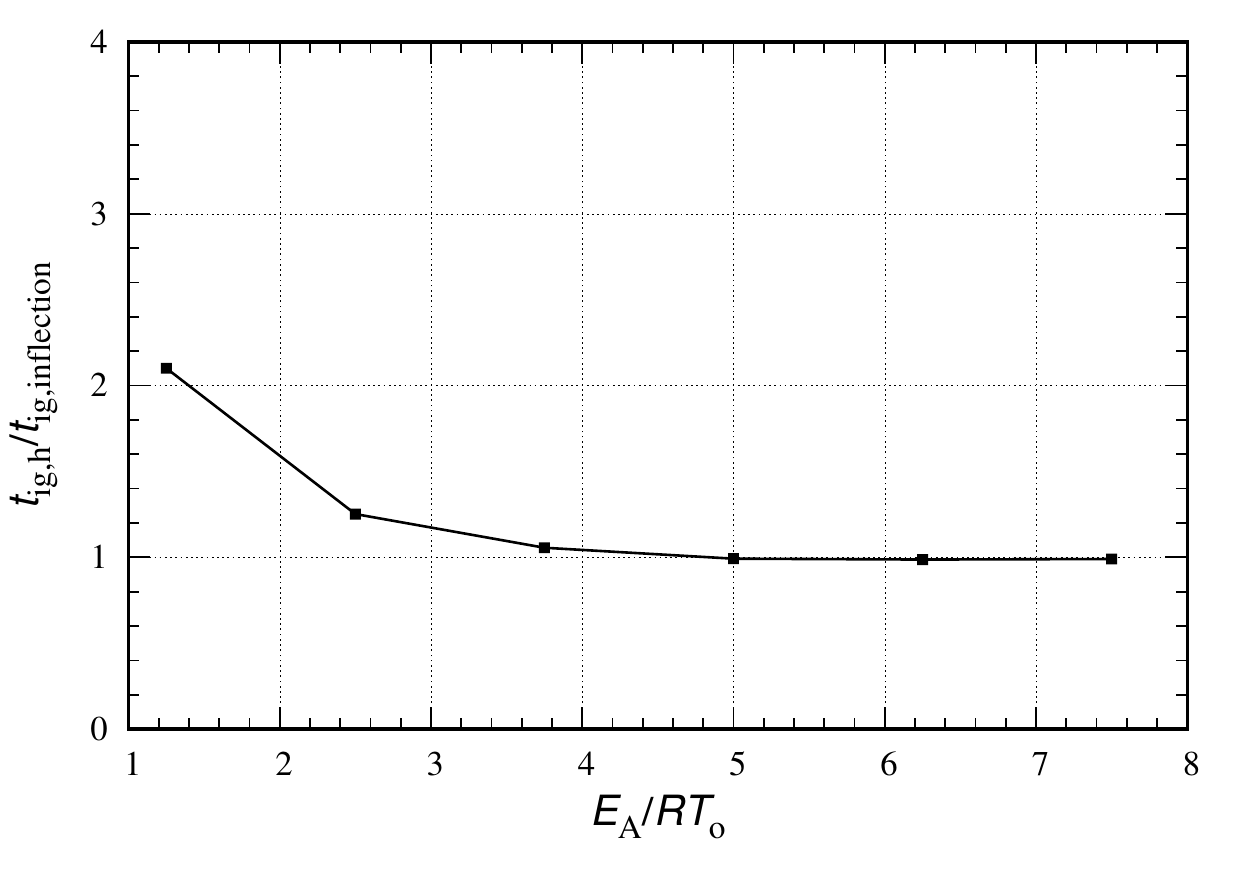}
	\caption{Ratio of igntion delay obtained at the depletion of half of the least abundant reactants to the ignition delay calculated at the point of inflection of the temperature profile for various $E_\text{A}/RT_\text{o}$ and $Q/RT_\text{o}$=2.5}
	\label{fig:tig_comparison}
\end{figure*}

For a given initial temperature and species concentrations, we are interested in the evolution of the system's temperature.  We define ignition as the time required for half of the energy of the system to be released, which corresponds to the depletion of half of the least abundant reactant. For reference, Fig.\ \ref{fig:tig_comparison} shows the ratio of ignition delay obtained by the above description to the ignition delay calculated at the point of inflection of the temperature profile. As expected \cite{doi:10.1080/00102207508946655}, both definitions of ignition delay are equivalent for high activation energies, typically above 5. 
\section{Numerical Method}
\label{Molecular Dynamics description}
The evolution of the particles positions and velocities follows the Event Driven Molecular Dynamics algorithm \cite{Alder_1959,Pochel_2005}.  The dynamics of hard sphere models can be determined analytically.  For any pair of particles, the collision time can be determined analytically as 
\begin{align}
t^*-t=-\frac{\vec{r}_{\text{ij}}\cdot \vec{v}_{\text{ij}}}{\vec{v}^2_{\text{ij}}}-\left[\bigg(\frac{\vec{r}_{\text{ij}}\cdot \vec{v}_{\text{ij}}}{\vec{v}^2_{\text{ij}}}\bigg)^2+\frac{R^2_{\text{ij}}-\vec{r}^2_{\text{ij}}}{\vec{v}^2_{\text{ij}}}\right]^{0.5}
\end{align}
where $t$ and $t^*$ are the current and next collision times, respectively; $\vec{r}_{\text{ij}}$, $\vec{v}_{\text{ij}}$ and  $R_{\text{ij}}=R_\text{i}+R_\text{j}$ are the relative distance, relative velocity and sum of the radius of the colliding spheres, respectively. The system is evolved from collision to collision, or event to event, hence the name of the algorithm. Our implementation of the method follows P$\ddot{\text{o}}$schel's procedure \cite{Pochel_2005}.  

\section{Simulation details}
\label{Simulation details}
\begin{figure*}
	\centering
	\subfloat[$t$ = $\tau_o$]{\includegraphics[width=67mm]{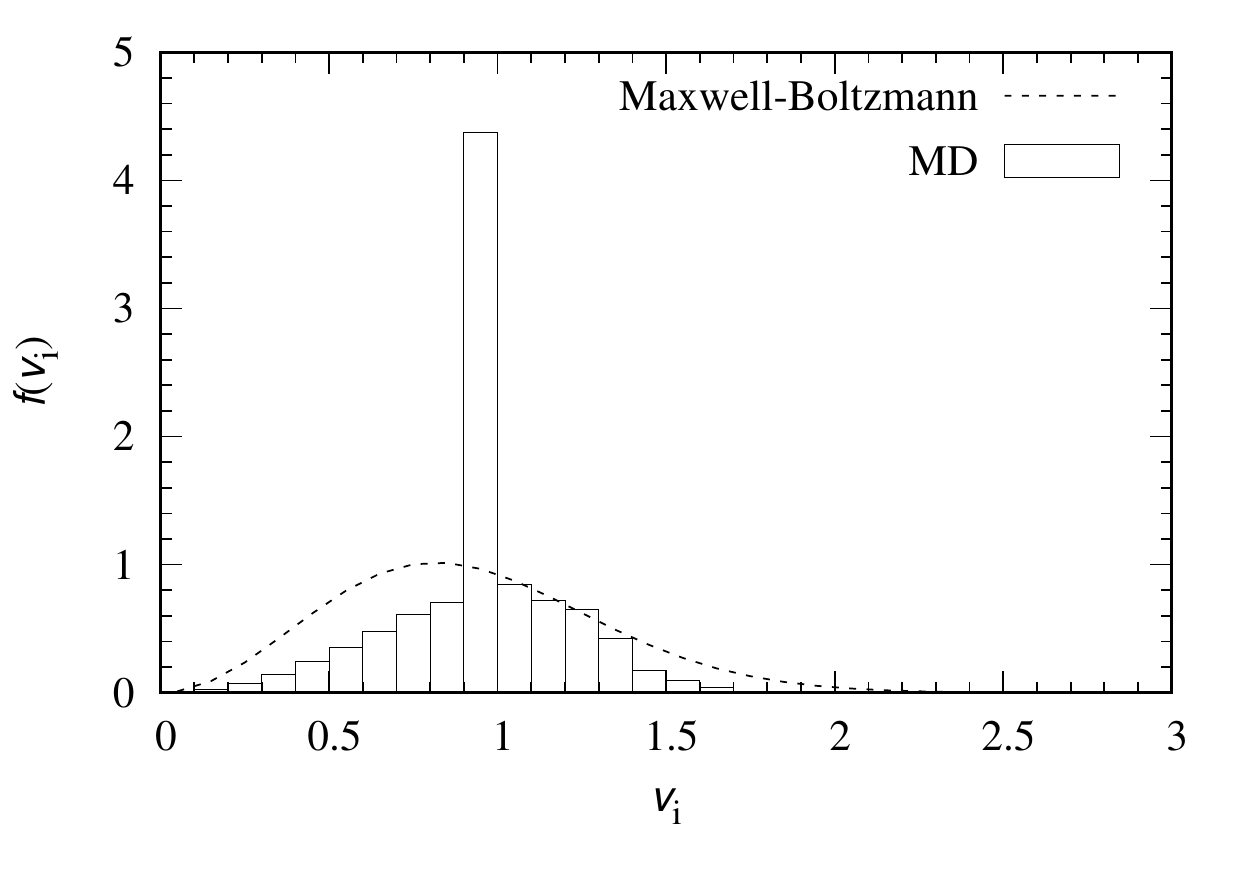}}\hfill
	\subfloat[$t$ = 3$\tau_o$]{\includegraphics[width=67mm]{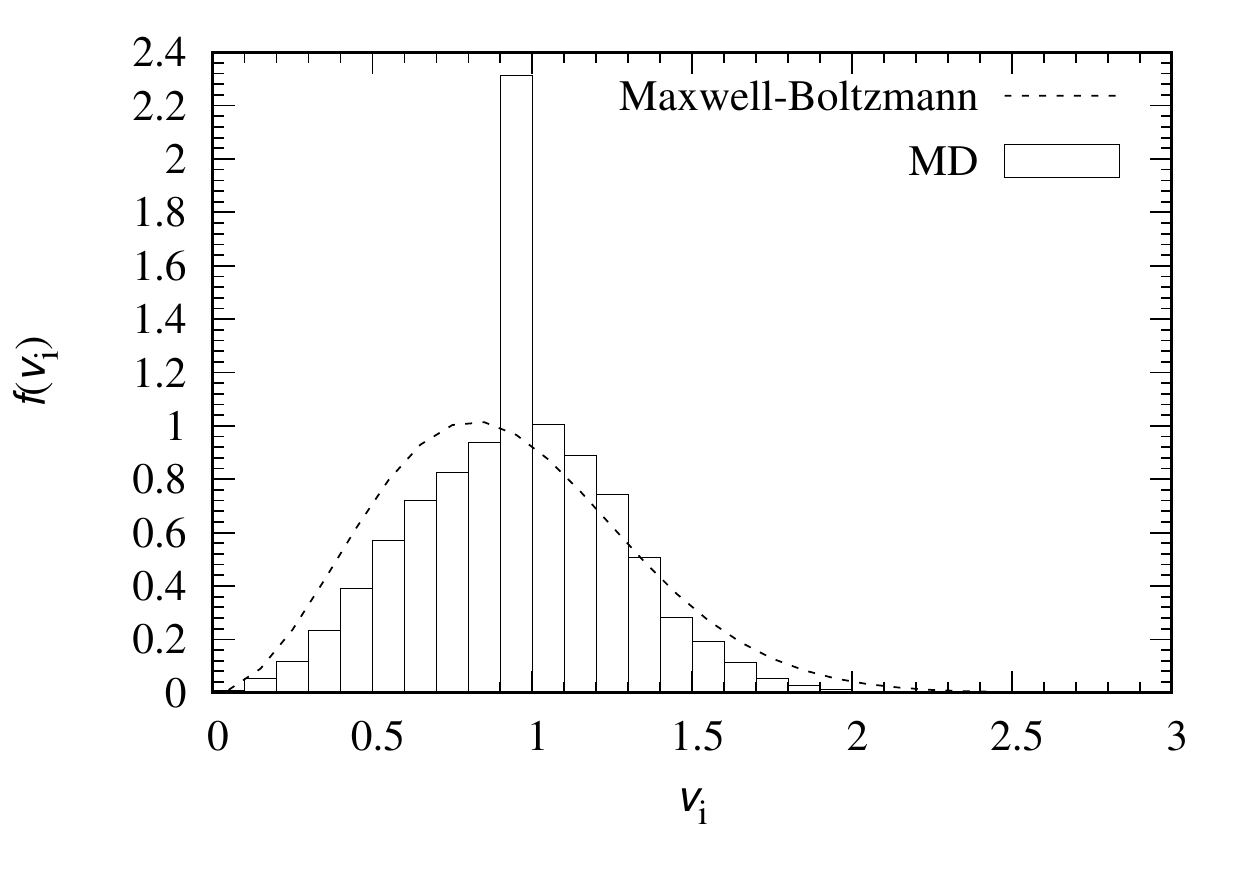}}\vfill
	\subfloat[$t$ = 5$\tau_o$]{\includegraphics[ width=67mm]{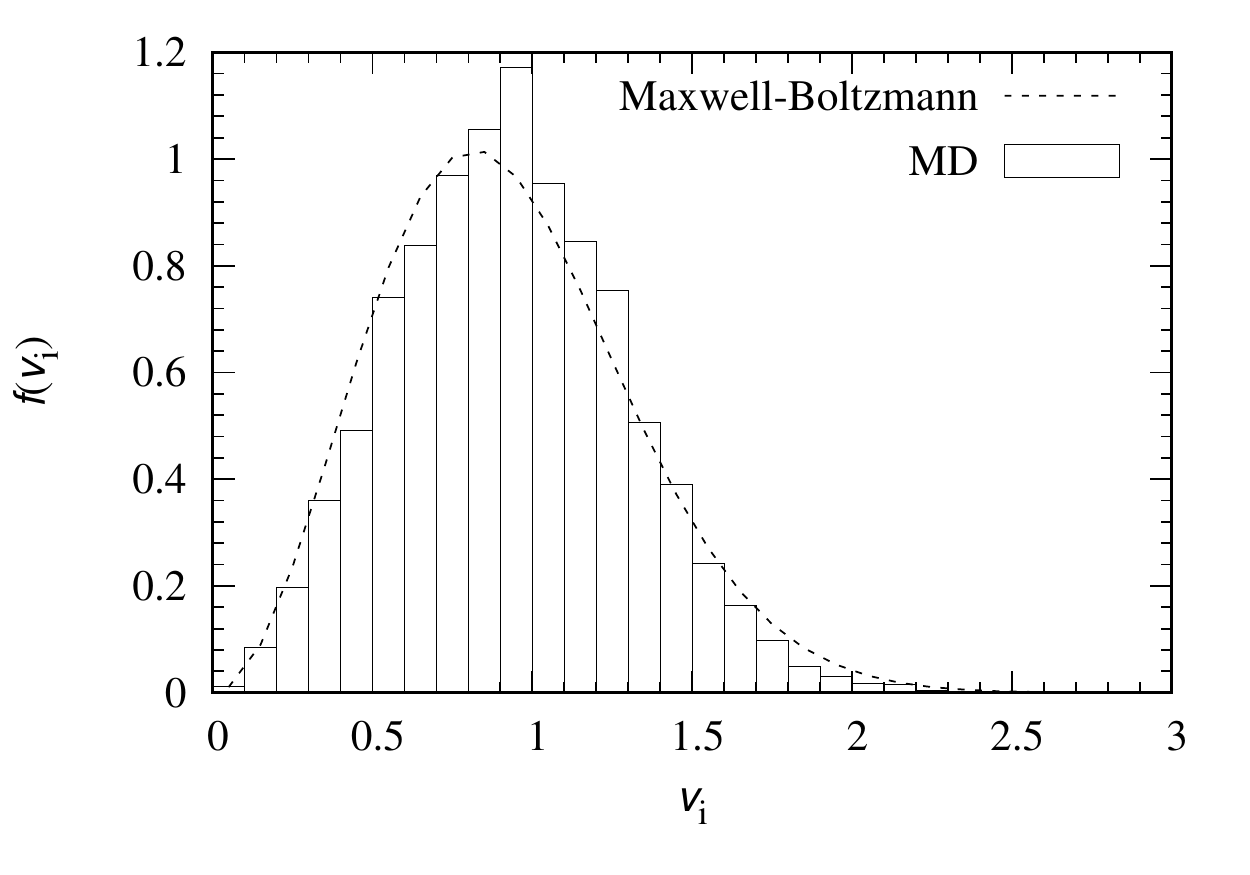}}\hfill
	\subfloat[$t$ = 10$\tau_o$]{\includegraphics[width=67mm]{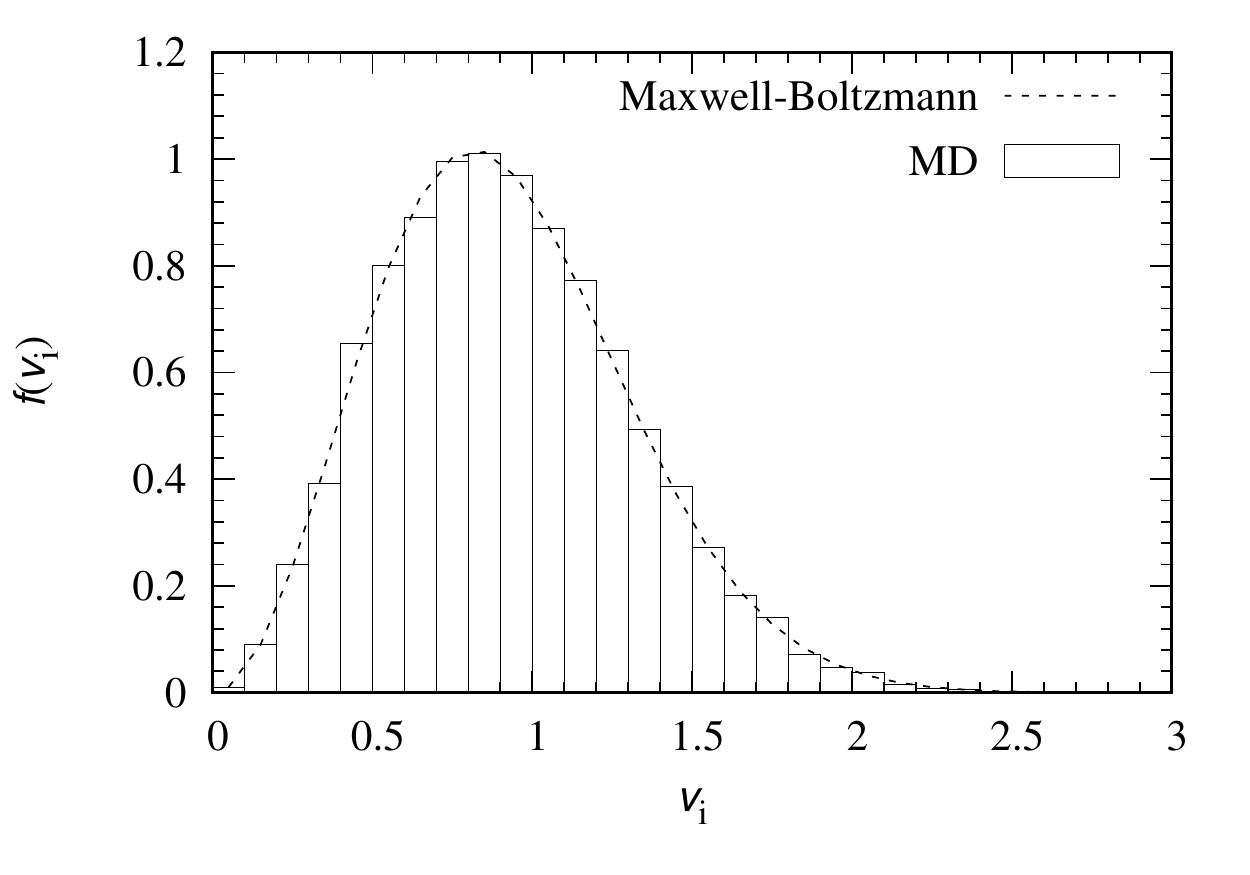}}
	\caption{Probability distribution of speeds obtained from MD for different point in time for $N$ = 10000 and compared with MB distribution.}
	\label{fig:MB distribution}
\end{figure*}

As initial condition, the number of type A and B spheres are specified and denoted as $N_A$ and $N_B$, respectively. In all calculations reported, unless otherwise stated, the number of particles of species A and B was identical. Each calculation was initialized by particles having equal speeds and randomized trajectories. This permits to accurately control the initial temperature (energy) of the system, as it is conserved exactly through elastic collisions. Collisions with the boundaries are considered as reflective. The system is then allowed to evolve with all reactive collisions turned off, until all particles undergo a sufficient number of collisions such that their speed distribution converges to the Maxwell-Boltzmann (MB) distribution, given by:
\begin{align}
f(v_\text{i})=\bigg(\frac{m}{2\pi kT}\bigg)^{3/2}\exp\bigg(-\frac{m{v_\text{i}}^2}{2kT}\bigg) 
\end{align} 
Figure \ref{fig:MB distribution} shows an example of the thermalization process. After one mean collision time (Fig. \ref{fig:MB distribution}a), there is a large peak in the distribution, corresponding to the initial speed of the particles. This peak progressively disappears as collisions deplete this class of particles; energy is eventually partitioned in classes of the tails of the distribution.   The  distribution of speeds is seen to evolve towards the MB distribution within less than 10 mean collision times, $\tau_o$.

\begin{figure*}
	\centering
	\includegraphics[trim={0.1cm 0.1cm 0.1cm 0.25cm},clip,width=.7\columnwidth]{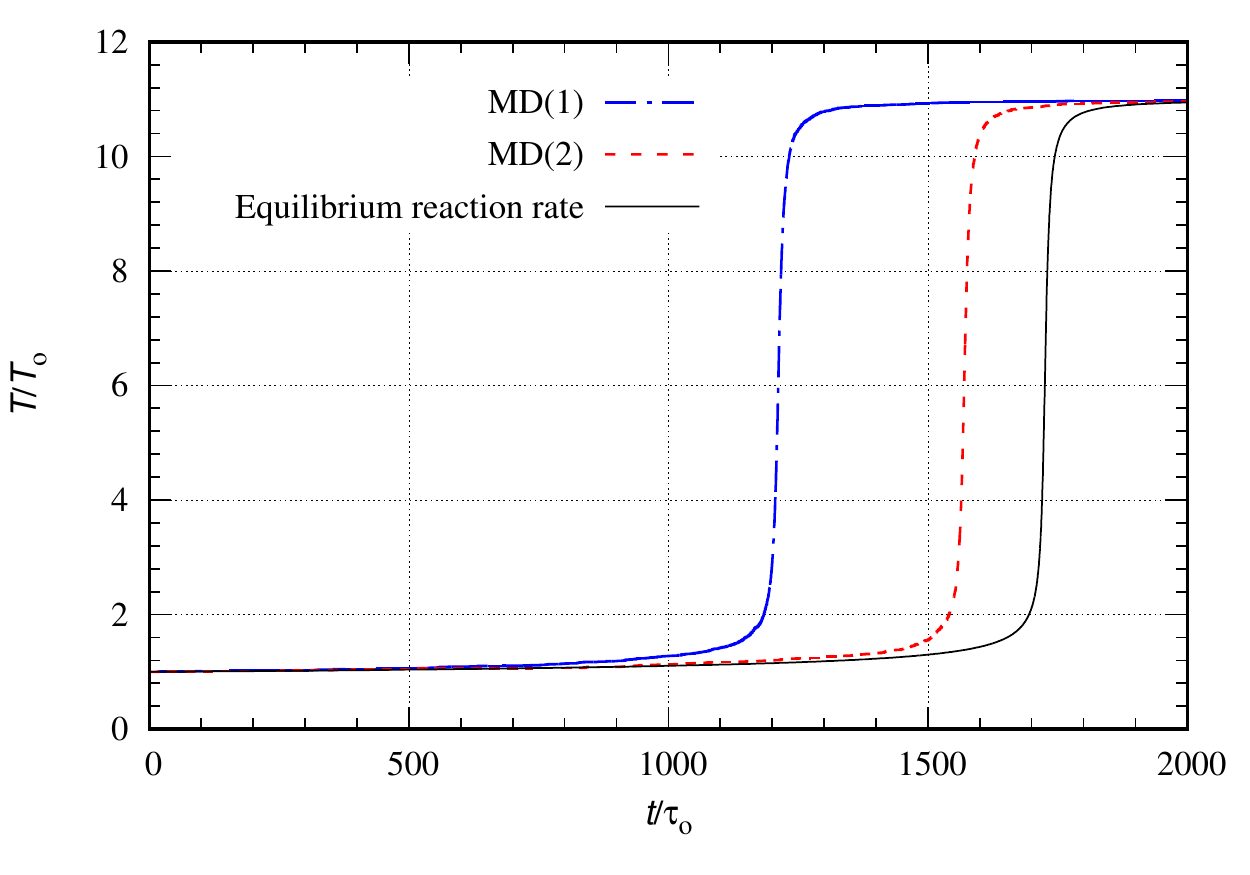}
	\caption{Example evolutions of temperature obtained for $N$=10000 with $Q/RT_\text{o}$=10 and $E_\text{A}/RT_\text{o}$=7.5.}
	\label{fig:temp-profile}
\end{figure*}

Once a thermal equilibrium is achieved among the reactants A and B, reactive collisions are allowed. This marks time 0 of the problem formulation. As very few particles have a high enough speed to give rise to a reactive collision, the rate of energy release in the box is initially slow. As time progresses and the energy released heats the system, more and more particles can undergo reactive collisions. This thermal ignition eventually raises the temperature of the system progressively faster, until depletion of the reactants moderates the rates. The temperature of the system is calculated from the root-mean-square speed of the spheres, $\bar{C}^2$, which is given by: 
\begin{align}
(\bar{C}^2)^{1/2} = (3RT)^{1/2}
\end{align}
Figure \ref{fig:temp-profile} shows the evolution of the system's temperature for two such calculations, which show the molecular noise, reflected by different configurations of the system at the same thermodynamic temperature give rise to a stochastic effect in the time of ignition.  We conducted the calculations 100 times for each set of parameters of interest, such that the ignition delay reported is the mean value. Likewise, the standard deviation is also recorded, as a measure of the stochasticity.
To test the influence of the system size, calculations have been performed by changing the total number of spheres, $N$, ranging from 50 to 10000, corresponding to cubes with sides, $L$, ranging from 1.103 to 6.45 mean free paths.

\begin{figure*}
	\centering
	\includegraphics[trim={0cm 0cm 0cm 0cm},clip,width=.7\columnwidth]{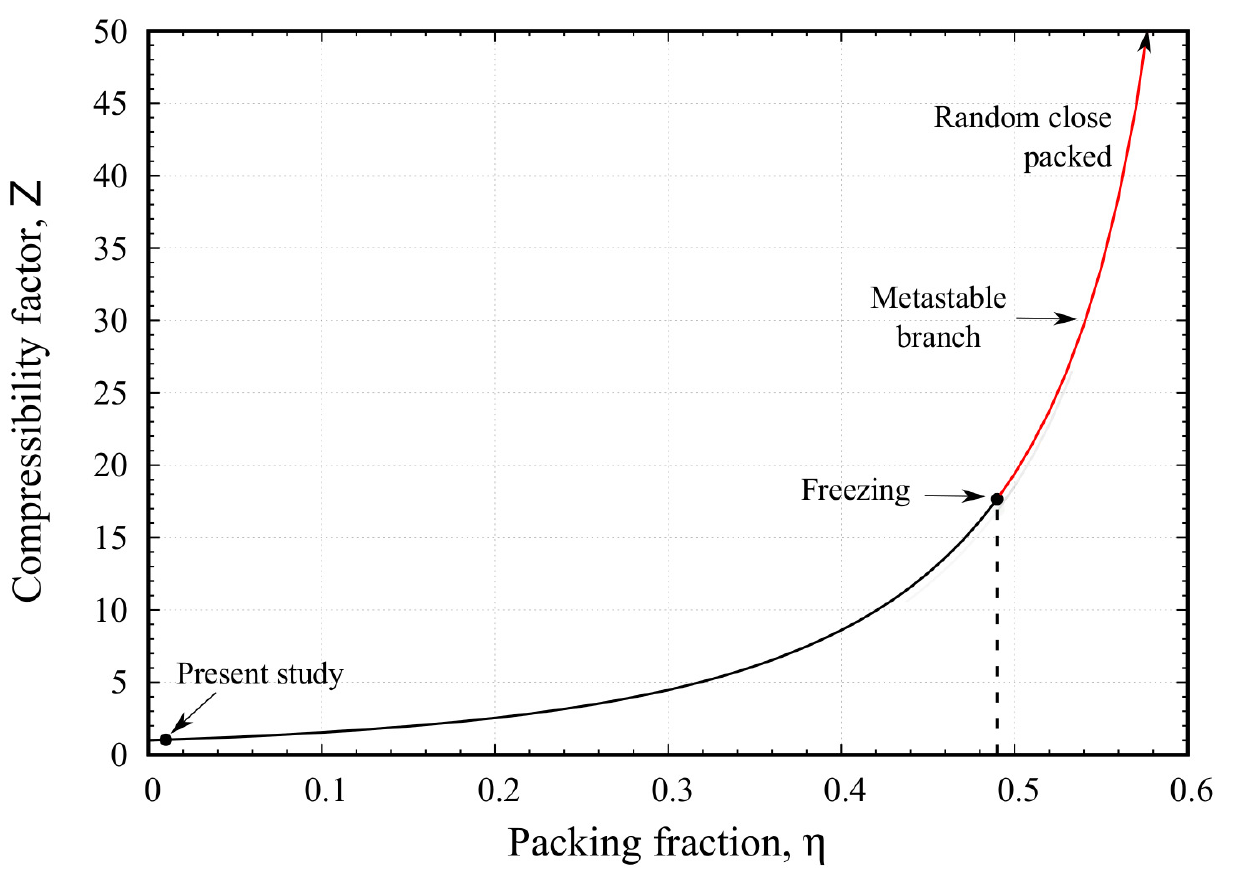}
	\caption{Phase diagram of an Hard sphere fluid \cite{Mulero_2008}}
	\label{fig:eta vs Z}
\end{figure*}

All calculations were performed in the ideal gas regime with a particle volume fraction of $\eta $ = 0.01.  For reference, Fig. \ref{fig:eta vs Z} shows the compressibility factor for various values of $\eta$ calculated from an equation of state for hard spheres provided by Song et al \cite{Song_1988}. The compressiblity factor for $\eta$ = 0.01 is found to be $Z$ = 1.04, which is close to unity (the ideal gas limit) and more than an order of magnitude away from the freezing phase transition. This emphasizes the fact that the value of $\eta$ used in the present work corresponds to a dilute, ideal gas regime.

\section{Continuum level description from kinetic theory}
\label{Continuum level description}
The results from molecular dynamics simulations described in the previous section were compared with those obtained from kinetic theory descriptions, where the reaction rate is either the standard form assuming thermal equilibrium at every instant, or its corrected versions to account for non-equilibrium effects in the evolution of the speed distribution function \cite{Prigogine_1949,Prigogine_1950}.  

For the problem of an insulated system at constant volume, its macroscopic evolution is given by \cite{Williams_1985}:
\begin{align}
\rho c_{\text{V}} \ode{T}{t}&=Q\omega_{\text{C}}\label{energy} \\  
\rho\ode{Y_{\text{C}}}{t}&=\omega_{\text{C}} \label{species}
\end{align}
where $Y_{\text{C}}$ and $\omega_{\text{C}}$ are the mass fraction and the production rate of product C, respectively; $c_\text{V}$ is the specific heat at constant volume.  Given an initial temperature $T_\text{o}$ and concentrations, the integration of these equations provide the evolution of the system's temperature and concentrations, and hence permits to determine the ignition delay.  

The standard rate of reaction, if one assumes a gas in local thermal equilibrium, takes the form \cite{Vincenti_1975}: 
\begin{align}
\omega_\text{C}=48\frac{\eta}{\sqrt{\pi}d}\rho Y_{\text{A}} Y_{\text{B}} \sqrt{RT} \exp\bigg(-\frac{E_\text{A}}{RT}\bigg) \label{reaction_homogeneous}
\end{align}
where $Y_{\text{A}}$ and $Y_{\text{B}}$ are the mass fractions of reactants A and B, respectively. $\eta$ is the volume fraction, $\rho$ is the density and $R$ is the gas constant. The pair correlation function g($\eta$) is 
\begin{align}
g_2(\eta) = \dfrac{(2-\eta)}{2(1-\eta)^3}
\end{align}

The initial mean free path and mean free time are , 
\begin{align*}
\lambda = \dfrac{\sqrt{\pi} d}{8\sqrt{3} \eta g_\text{2}(\eta)}; ~~~~~
\tau_\text{o}=\lambda/u_{\text{rms(o)}} 
\end{align*}
\\
For the current study the length and the time scales are normalized by initial mean free path and initial mean free time of the gas. With this scaling the homogeneous ignition description is independent of $\eta$, allowing $Q/RT_{\text{o}}$ and $E_{\text{A}}/RT_{\text{o}}$ to uniquely define the system's evolution.

\section{Results and discussion}
\label{Results}
\begin{figure*}
	\centering
	\includegraphics[trim={0.1cm 0.1cm 0.5cm 0.25cm},clip,width=.7\columnwidth]{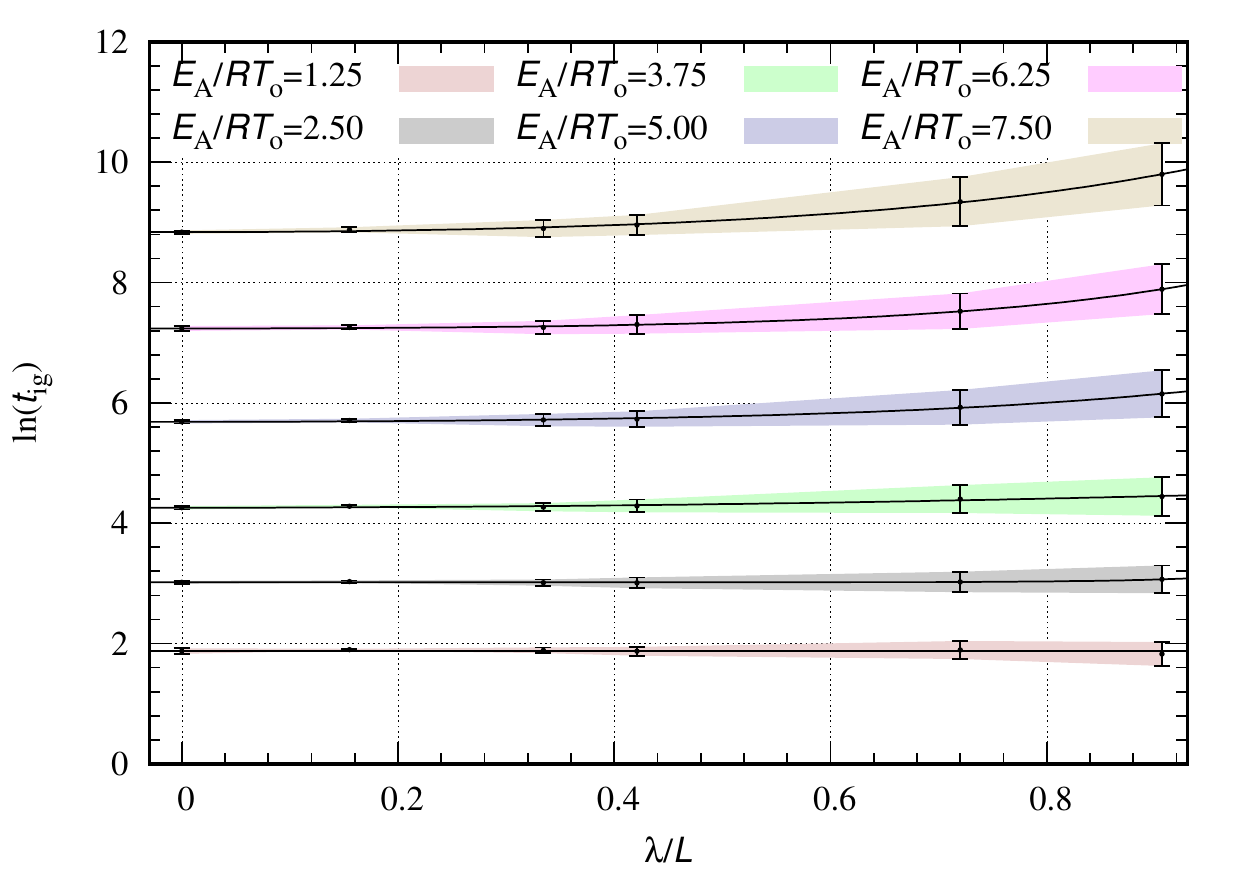}
	\caption{Logarithm of ignition delay extrapolated to infinite domain size $Q/RT_{\text{o}}$ = 2.5. The error bars  indicate the standard deviation of ignition delay time. The color band represents the confidence zone of MD results for each activation energy. Solid line indicates the asymptote calculated from Eq. \eqref{asymptote} }
	\label{fig:domainsize_Q2.5}
\end{figure*}

\begin{figure*}
	\centering
	\includegraphics[trim={0.1cm 0.1cm 0.5cm 0.25cm},clip,width=.7\columnwidth]{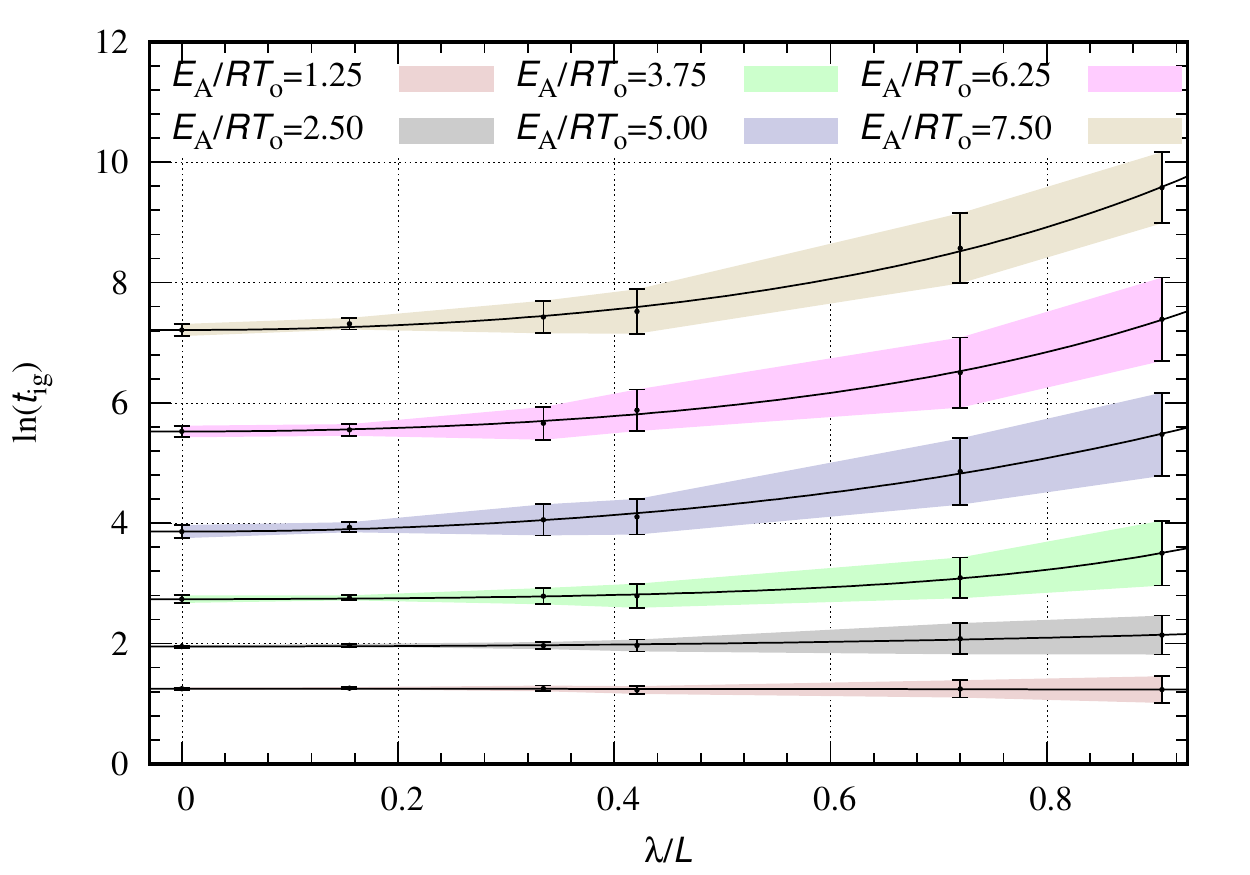}
	\caption{Logarithm of ignition delay extrapolated to infinite domain size for $Q/RT_{\text{o}}$ = 10. Same description as Fig. \ref{fig:domainsize_Q2.5} }
	\label{fig:domainsize_Q10}
\end{figure*}

The results for the ignition delay determined using the MD model described above are shown in Figs.\ \ref{fig:domainsize_Q2.5} and \ref{fig:domainsize_Q10} for the two values of heat release considered, respectively $Q/RT_{\text{o}}$=2.5 and 10. Each data point represents the average ignition delay among 100 calculations, while the error bar is the standard deviation. Each color band in the figure indicates the confidence zone of MD results for a particular activation energy for varying domain size. For all values of $Q$ and $E_{\text{A}}$, the calculated asymptote falls within the confidence zone of ignition delay time. It is prominent from the figure that the ignition delay and standard deviation decrease with increasing domain size. For low activation energies, regardless of heat release, the ignition delay time is very weakly influenced by the domain size effects. However, for increasing $Q$ and $E_{\text{A}}$, the trend line is steeper, showing a stronger influence of domain size. For all the calculations performed, the ignition delay was found to converge well to a system of infinite size (abscissa of 0 in the figures).  

In order to determine the extrapolated value of ignition delay in an infinite system, the ignition delay dependence on system size was assumed to have the form 
\begin{figure*}
	\centering
	\includegraphics[trim={0cm 0.25cm 0.5cm 0.25cm},clip,width=.7\columnwidth]{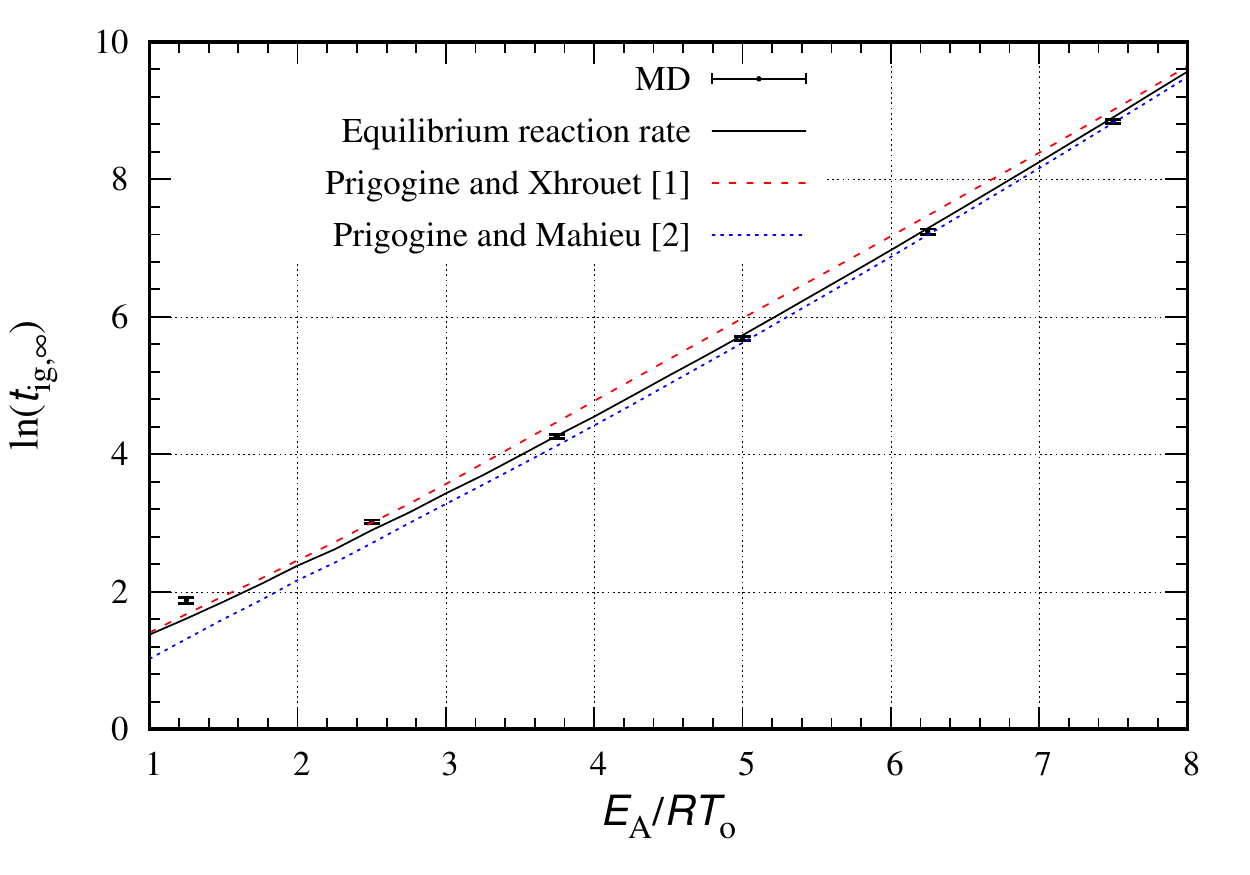}
	\caption{ Ignition delay time with varying $E_{\text{A}}$ obtained for $Q/RT_{\text{o}}$ = 2.5 from MD (errorbars) are compared with equilibrium reaction rate (solid line) and non-equilibrium reaction rate calculated from Eqs. \eqref{correction_Xhrouet} (dashed red line) and \eqref{correction_Mahieu} (dotted blue line)}
	\label{fig:comp-inf_domain_Q2.5}
\end{figure*}
\begin{figure*}
	\centering
	\includegraphics[trim={0cm 0.25cm 0.5cm 0.25cm},clip,width=.7\columnwidth]{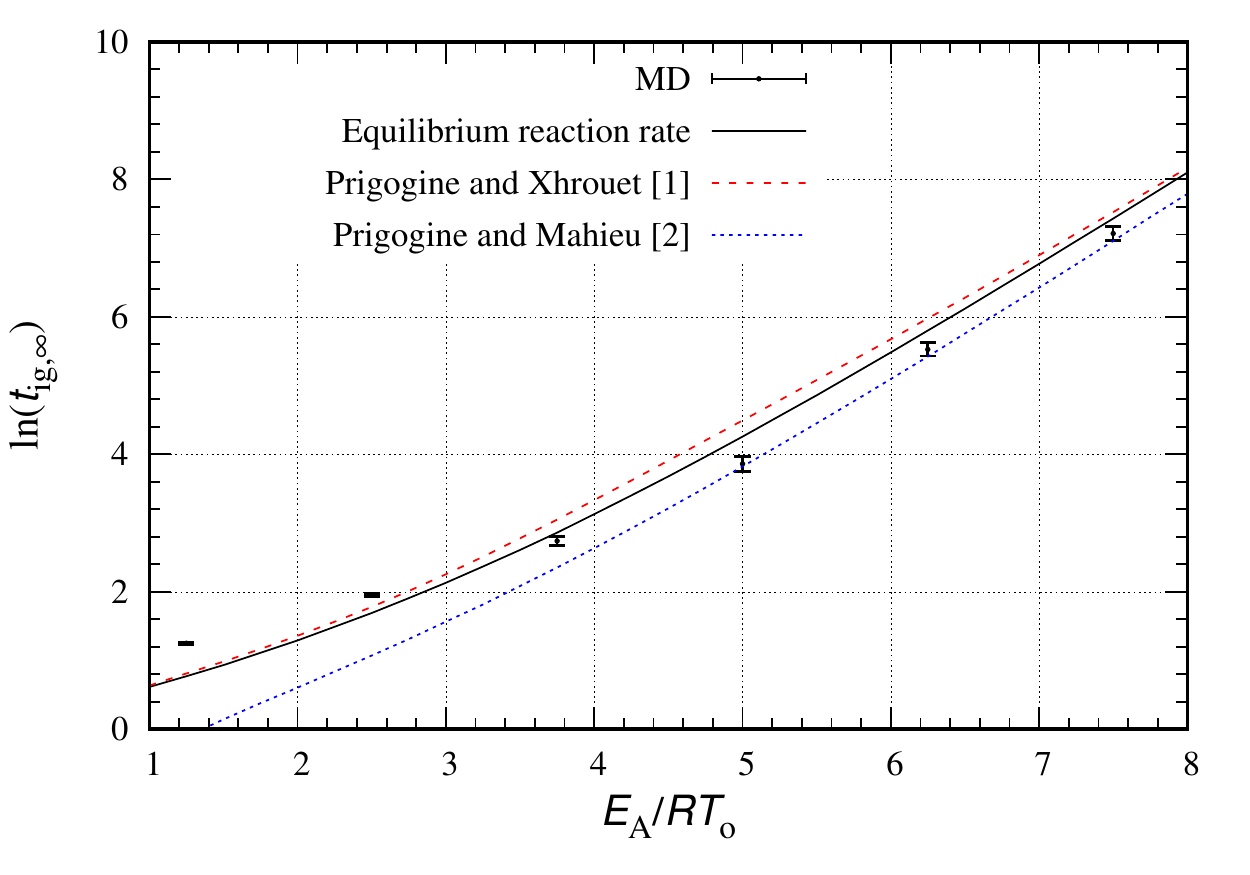}
	\caption{Ignition delay time with varying $E_{\text{A}}$ obtained for $Q/RT_{\text{o}}$ = 10. Same description as Fig.  \ref{fig:comp-inf_domain_Q10}}
	\label{fig:comp-inf_domain_Q10}
\end{figure*}
\begin{align}
\ln(t_{\text{ig}}) = \ln(t_{\text{ig,$\infty$}}) + b \cosh\bigg(\frac{c\lambda}{L}\bigg)
\label{asymptote}
\end{align}
where $b$, $c$ and $t_{\text{ig,$\infty$}}$ are fitting constants, the latter being the desired ignition delay time at infinite domain size. The fits are also shown in the Figs. \ref{fig:domainsize_Q2.5} and \ref{fig:domainsize_Q10}. The presence of uncertainty is inherent in almost all extrapolation methods, which we also represented by error bar for the extrapolated values in Figs \ref{fig:domainsize_Q2.5} and \ref{fig:domainsize_Q10}. Therefore, the obtained values are valid with a confidence of approximately 68\%. 

The ignition delays extrapolated to infinite domains are shown in terms of the activation energy in Figs. \ref{fig:comp-inf_domain_Q2.5} and \ref{fig:comp-inf_domain_Q10} for $Q/RT_{\text{o}}$=2.5 and 10, respectively. These are compared with three predictions of ignition delay calculated with expressions for the reaction rate from kinetic theory:  the thermal equilibrium standard reaction rate, equation \eqref{reaction_homogeneous}, the rate perturbed by Prigogine and Xhrouet for an isothermal system and the rate perturbed by Prigogine and Mahieu incorporating the effect of energy release. 

For the isothermal system,  Prigogine and Xhrouet derived the rate $\omega_{\text{X}}$ \cite{Prigogine_1949}: 

\begin{align}
\omega_{\text{X}}=\omega_C \bigg\{ 1-\frac{1}{32}e^{-\frac{E_{\text{A}}}{RT}} \bigg(\frac{E_{\text{A}}}{RT} \bigg)^3 \bigg[ \bigg( \frac{E_{\text{A}}}{RT} \bigg)^2-5\frac{E_{\text{A}}}{RT}+\frac{17}{2} \bigg]\bigg \} \label{correction_Xhrouet}
\end{align}

Incorporating the effects of reaction, the non-equilibrium reaction rate, $\omega_{\text{M}}$, derived by Prigogine and Mahieu is \cite{Prigogine_1950}, 
\begin{align}
\omega_{\text{M}} = \omega_C \bigg\{1+1.2 Y_{\text{A}} Y_{\text{B}} \bigg(\frac{Q}{E_{\text{A}}}\bigg)\bigg\} \label{correction_Mahieu}
\end{align}
\begin{figure}
	\centering
	\includegraphics[trim={0cm 0.25cm 0.5cm 0.25cm},clip,width=.7\columnwidth]{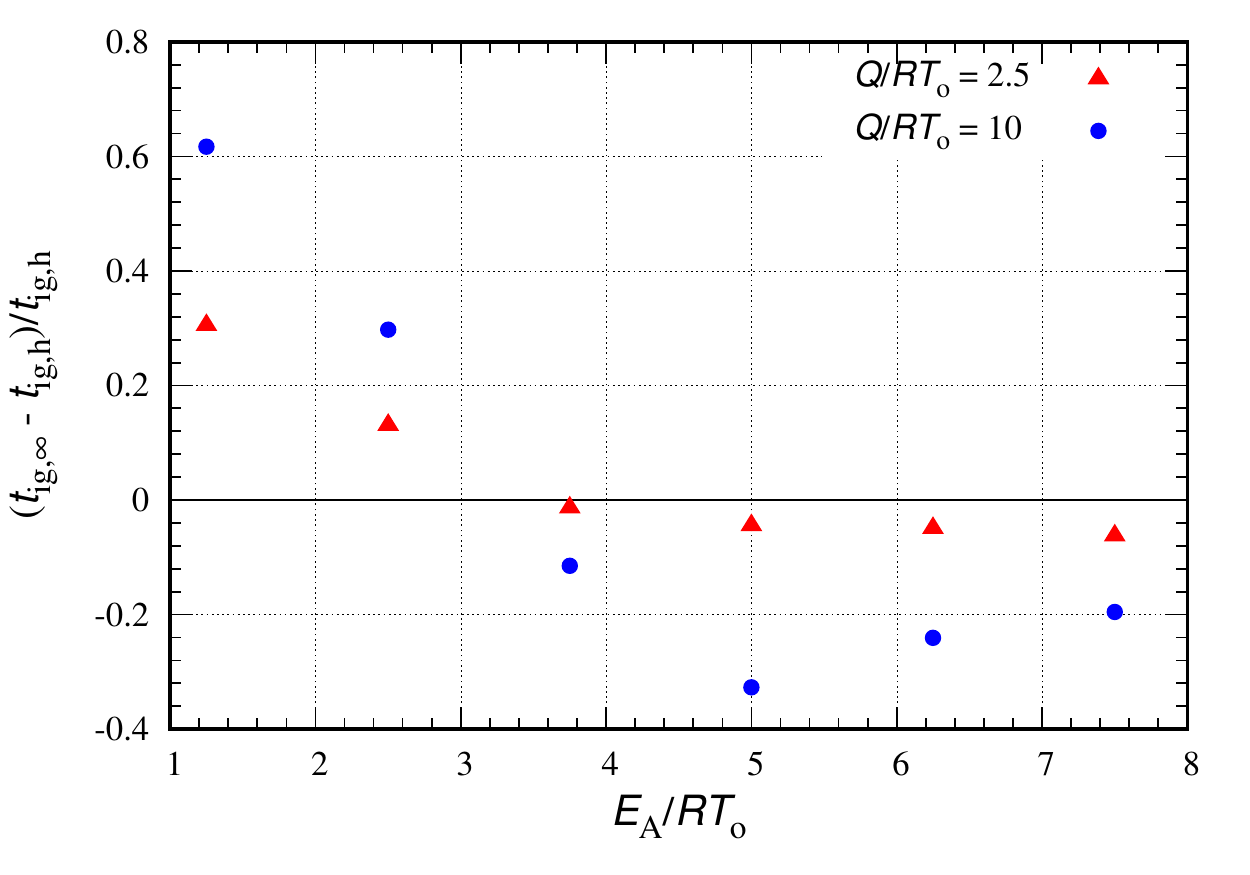}
	\caption{Relative difference between ignition delay time obtained from MD and homogeneous mixture.}
	\label{fig:tig-tigh_reldiff-plot-inf}
\end{figure}
The reaction rate obtained from the first model, equation \eqref{correction_Xhrouet}, is found to be lower than the standard reaction rate by 9\%, 23\% and 20\% for activation energies of 2.5, 5 and 7.5, respectively. Similarly, for the second model, the reaction rate calculated from equation \eqref{correction_Mahieu} predicts a reaction rate higher than the equilibrium reaction rate by 10\%, 30\%, 40\% and 120\% for $Q/E_{\text{A}}$ ratio of 0.3, 1, 1.3 and 4, respectively. This explains the fact of longer ignition delay predicted by the first model and the shorter ignition delay predicted by the second model. The comparison of our molecular dynamics results for ignition delay in Figs.\ \ref{fig:comp-inf_domain_Q2.5} and \ref{fig:comp-inf_domain_Q10} with those obtained from kinetic theory permits to establish certain clear trends.  First, we would like to compare with the results of standard kinetic theory.  The differences from the standard kinetic theory are also shown in Fig.\ \ref{fig:tig-tigh_reldiff-plot-inf}. For the low value of energy release of $Q/RT_{\text{o}}$=2.5, the standard equilibrium kinetic theory captures well, although slightly over-predicts the ignition delay obtained from the molecular dynamics calculations.  Larger deviations are observed for low activation energies, below $E_{\text{A}}/RT_{\text{o}}$ of approximately 3.  In this regime, the ignition delay was found to be longer than predicted by as much as 30\% at low activation energies.  For the high energy release of $Q/RT_{\text{o}}$=10 shown in Fig.\ \ref{fig:comp-inf_domain_Q10}, the predictions of the standard kinetic rate is significantly deteriorated for all activation energies.  The largest departure is seen for intermediate activation energies of $E_{\text{A}}/RT_{\text{o}}$ = 5, where the ignition delay calculated from molecular dynamics is more than 30\% shorter than anticipated from the standard equilibrium rate.  Likewise, for low values of activation energy,  the equilibrium model predicts an ignition delay longer by 60\%. 

The isothermal model prediction of Prigogine and Xhrouet, shown in red, systematically predicts an ignition delay longer than the one obtained from the equilibrium value.  Nevertheless, it does not capture any of our molecular dynamic results.   At large activation energies, this is the opposite trend from both our low heat release and high heat release results.  At low activation energies, it provides a correction in the correct direction of lengthening the induction delay, but fails to capture the trend obtained with activation energies approaching 1. According to Prigogine et al., at high-temperature, the fraction of reactive collisions is sufficiently high even during the early phases of the run-away process \cite{Prigogine_1949}. Therefore, they proposed that the departure from equilibrium can be attributed to the reactions occurring at a faster rate than they are able to transfer their kinetic energy to the non-reacted spheres. Thus, moving the system away from equilibrium. It is these non-equilibrium effects caused by the  reactions that yield a higher ignition delay.

Interestingly, the model of Prigogine and Mahieu provides a much better approximation to both of our low heat release and high heat release results for sufficiently high activation energies only \cite{Prigogine_1950}. This model includes the importance of heat of reaction, which can perturb the Maxwell distribution to an appreciable extent in the order of $Q/E_{\text{A}}$. As a result, the rate of reaction is increased for exothermic reactions and decreased for endothermic reactions. Also, this model demonstrates the inclusion of a sufficiently large heat of reaction which yield a non-equilibrium reaction rate larger than the one derived with the assumption of local equilibrium, using the same perturbation method. Therefore this model is found to be better than the former for our particular problem of thermal ignition, where exothermicity plays a strong role in introducing non-equilibrium effects and modifying the macroscopic rate of reactions. For $E_{\text{A}}/RT_{\text{o}}$ greater than 5, the prediction is within the standard deviation associated with the stochasticity of the ignition.  This result is encouraging, as the model captures the data at the linear order correction in $Q/E_{\text{A}}$, not requiring higher order terms in the Chapman-Enskog perturbation.  At lower activation energies, however, the prediction significantly diverges from the molecular dynamic data for both low and high heat release. 

\begin{figure}
 	\centering
 	\includegraphics[trim={0cm 0.25cm 0.5cm 0.25cm},clip,width=.7\columnwidth]{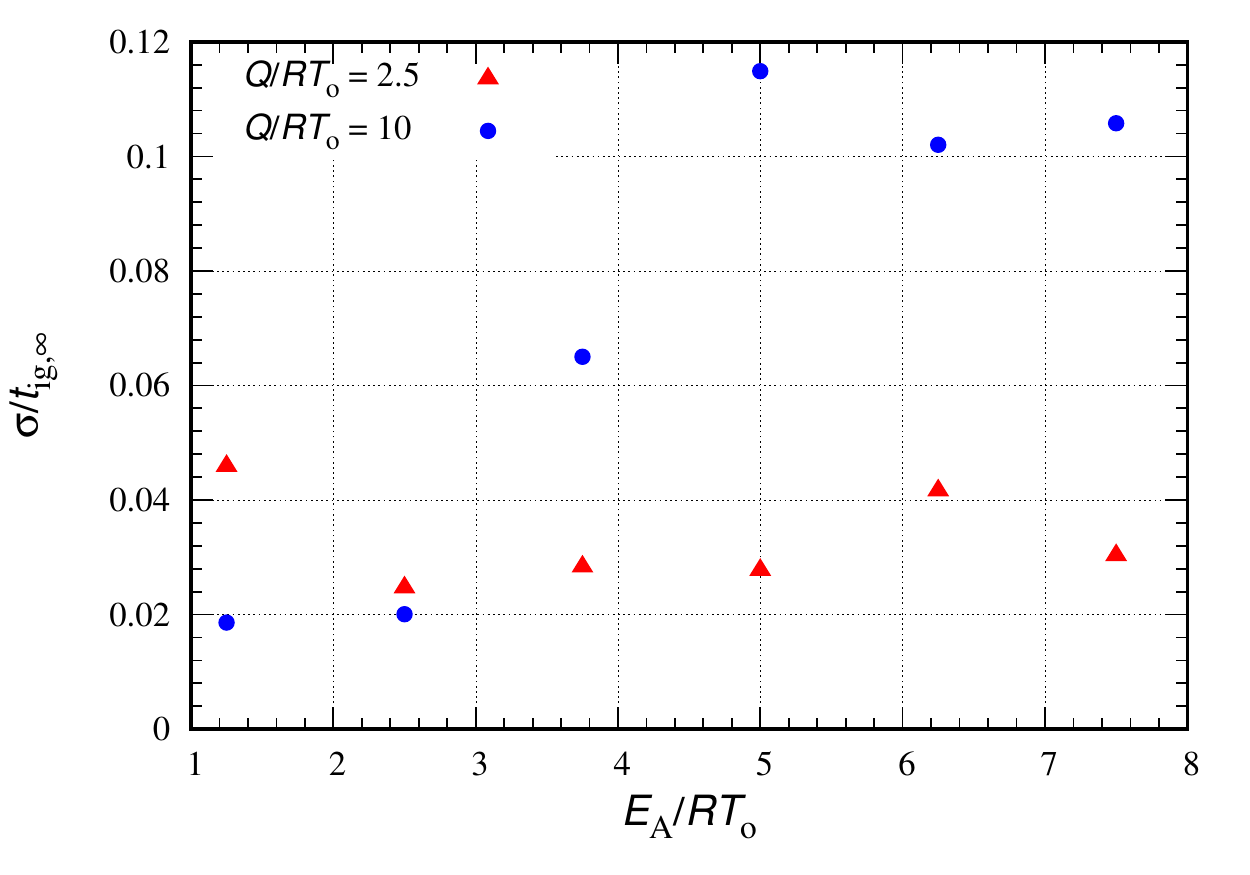}
 	\caption{Standard deviation of ignition delay times obtained from MD for different values of heat release}
 	\label{fig:tig-tigh_std-plot_inf}
 \end{figure}

At low temperatures (high activation energy), the good agreement between our calculations and those using the perturbed rate of Prigogine and Mahieu supports the validity of the physical mechanism proposed by them. Highly energetic particles resulting from reactive collisions promote local reactions.  These do not have enough time to equilibrate with the rest of the system, thus giving rise to what we would like to call \textit{molecular hotspots}.  This result is also in accordance with the previous numerical results by Sirmas and Radulescu in the two-dimensional system \cite{Sirmas_2017}.  In this regime of ignition, where non-equilibrium effects significantly promote ignition, we also observe a much higher stochasticity of ignition, as shown in Fig.\ \ref{fig:tig-tigh_std-plot_inf}.  As the heat release and activation energy are increased, so is the stochasticity among our 100 
simulations, measured by the standard deviation.  Interestingly, this is also the regime where ignition is \textit{mild} in practice: it takes the form of localized ignition spots \cite{Meyer_1971, Gardner_2010}. In this regime, the picture that emerges is the ever increasing role of molecular fluctuations. 

In our previous study, we compared the results only with reaction rate assuming homogenous ignition \cite{Sirmas_2017}. Whereas, in the present work along with the equilibrium reaction rate, we compared our results with the existing kinetic theory models which takes into account the effect of heat of reaction while calculating the reaction rate. For high $Q$ and low $E_\text{A}$, the ignition delay is found to be 50\% longer in 2D whereas, it is 60\% in 3D. Similarly, for high $Q$ and high $E_\text{A}$, the delay time is approximately 30\% lower in both 2D and 3D. However, the low heat release case is not much affected by the presence of third dimension and recovers the 2D results well by roughly about 3\% difference. When compared to 2D simulations, the standard deviation of ignition delay for both low and high heat release case is reduced almost by 2\% and 20\%, respectively. These results show that the statistical fluctuations become important with growing $Q$ and $E_\text{A}$. Also, it is to be noted that the ignition delay calculated in 2D are not extrapolated to infinite domain as discussed in this study. Hence, we can conclude that the magnitude of the results is different, but the trend followed by each set of parameter recovers the same as obtained from 2D.

For all the simulations discussed above, the proportion of species A and B were equal and no diluent was introduced.  However, in actual combustion systems, other species may be present, hence representing a potential heat bath to absorb part of the energy released and diminish the magnitude of the non-equilibrium effects documented.   We have addressed this effect of dilution in separate calculations in an initial mixture of 9A+B, where the excess A essentially represent a diluent.  Table \ref{table1:Inert simulations} reports the results performed in the undiluted and diluted cases for $Q/RT_{\text{o}}$ = 10 and $E_\text{A}/RT_{\text{o}}$ = 7.5 using $N$ = 10000 particles. In the diluted regime, the ignition delay was found lower than the one predicted with the assumption of Maxwell-Boltzmann distribution by approximately 5\%, whereas the difference was approximately 11\% for the non-diluted case. This clearly shows that non-equilibrium effect reported will be a strong function of dilution.  These are left for future study.

\begin{table}
	\centering
	\caption{Comparison of ignition delay time from MD for $N_\text{A}$ = 9 $N_\text{B}$ with that obtained for homogeneous ignition for $E_{\text{A}}/RT_{\text{o}}$=7.5 and $Q/RT_{\text{o}}$ =10 }
	\label{table1:Inert simulations}
	\begin{tabular}{c c c c c c c}
	\hline
	&\multicolumn{6}{c}{Ignition time ($t_{\text{ig}}$), $N$ = 10000} \\
	\hline
	&\multicolumn{3}{c}{A+B} & \multicolumn{3}{c}{9A+B} \\ \hline
	&\multirow{2}{*}{Equilibrium} & \multirow{2}{*}{\shortstack{Molecular \\ Dynamics}} & \multirow{2}{*}{\shortstack{Standard \\ Deviation}}& \multirow{2}{*}{Equilibrium}& \multirow{2}{*}{\shortstack{Molecular \\ Dynamics}} & \multirow{2}{*}{\shortstack{Standard \\ Deviation}} \\ \\ \hline
	&1682.4 & 1502.6 & 151.5 & 4868.4 & 4631.5 & 403.1 \\ 
	\multirow{2}{*}{\shortstack{Relative \\Difference}} &\multicolumn{3}{c}{\multirow{2}{*}{-10.98}} &\multicolumn{3}{c}{\multirow{2}{*}{-4.86}} \\ \\
	\hline
\end{tabular}
\end{table}
\begin{table}	
	\caption{Activation temperature $T_a$(K), heat of reaction $\bar{Q}$(J/mol) and heat of reaction temperature $\bar{Q}/\sum_{Reactants} \nu_i R_u$ (K) of reactions controlling ignition processes in hydrogen (partially adapted from Sanchez and Williams \cite{Sanchez_2014}- see text).}
	\label{table3: reduced hydrogen chemistry}
	{
		\centering
		\begin{tabular}{ c l r r r} 	
			\hline
			& &	 $T_a$(K) & $\bar{Q}$(J/mol) & $\bar{Q}/\sum_{Reactants} \nu_i R_u$(K)\\ 
			\hline
			1 &\ce{H + O_2 $\rightarrow$ OH + O} & 8590 & -70158 & -4219   \\ 
			2&\ce{H_2 + O $\rightarrow$ OH + H} & 3165 & -7810 &  -470  \\ 
			3&\ce{H_2 + OH $\rightarrow$ H_2O + H} & 1825 & 62814 & 3777   \\ 
			4&\ce{H + O_2 + M $\rightarrow$ HO_2 + M} & 0.0  & 196599 & 11823  \\ 
			5&\ce{H2 + O2 $\rightarrow$ HO2 + H} & 27888 & -239399 & -14397   \\ 
			6&\ce{H2O2 + M $\rightarrow$ OH + OH + M} & 25703 & -213974 & -25736  \\ 
			7&\ce{HO2 + HO2 $\rightarrow$ H2O2 + O2}& 5556 & 179110 & 10771   \\ 
			8&\ce{HO2 + H2 $\rightarrow$ H2O2 + H}& 12045 & -60289 & -3625  \\  
		\hline
	\end{tabular}\\~\\
	}  
\end{table}

It is also of interest to comment on the relevance of the present results for problems of auto-ignition in combustion systems, as for example the oxidation of hydrogen by oxygen or air.  It is now relatively well established that low temperature ignition in hydrogen is governed by a thermal mechanism of initiation \cite{Sanchez_2014}.  Nevertheless, the reaction network controlling the ignition process is not a single reaction, but a local network of reactions.  In hydrogen chemistry, the chemical mechanism is sufficiently simple to qualitatively comment on the potential role of non-equilibrium effects discussed in the present paper.  The eight important reactions are listed in Table \ref{table3: reduced hydrogen chemistry} \cite{Sanchez_2014}.  Also listed in this table is the activation temperature and an equivalent heat of reaction we calculated in dimensions of Kelvin ($Q/R$), such that their division by the local temperature yields the relevant parameters $E_a/RT = T_a/T$ and $Q/RT$.  For reference, the relevant low temperature regime of auto-ignition that we wish to discuss is around 1000 K - 1300 K depending on the pressure \cite{Sanchez_2014}.  This shows that non-dimensional activation energies lie in the range 0 to 30 and the heat release in the range -26 (negative means endothermic) to 12.  This shows that non-equilibrium effects associated with low activation temperatures discussed in the present study may directly affect the reaction rates of reactions 2, 3, 4 and 7.   On the other hand, the hot-spot mechanism associated with high heat release and high activation energy may \textit{directly} affect mostly reaction 7, as other highly energetic recombination reactions (e.g., reaction 4) have negligible activation.  Likewise, reactions 5 and 6 may be slowed down through equivalent cold spots due to their strong endothermicity.  

Nevertheless, reaction networks may bring non-equilibrium hot spot effects in sequence of energetic reactions (giving rise to non-equilibrium high speed molecules \textit{B}) followed by high activation reactions involving the molecule \textit{B}.  This is precisely the case for hydrogen ignition at low temperatures.  At temperatures below cross-over, hydrogen auto-ignition is controlled mainly by the auto-catalytic role of hydrogen peroxide (H$_2$O$_2$) through the sequence of reactions 6, 3, 4 and 7, leading to the global reaction 2H$_2$+O$_2$ $\rightarrow$ 2H$_2$O, with the rate controlling step given by reaction  6  \cite{Sanchez_2014} .   Reaction 6 has a very high activation energy, and produces two OH.  These proceed through reaction 3, which occurs twice and produces two H atoms which are then consumed by reaction 4 to form two HO$_2$ in an energetic reaction.  These two energetic HO$_2$ molecules then form a H$_2$O$_2$ molecule in the activated and highly energetic reaction 7 to complete the auto-catalytic loop, which occurs locally with negligible influence by transport phenomena.  Of interest in this loop is that the highly energetic hydrogen peroxide molecule then restarts the loop in a very high activation energy reaction.  The auto-catalytic H$_2$O$_2$ loop thus has the form of the model reaction studied in the present study.  Highly energetic products of an exothermic reaction promote local temperature non-equilibrium, which then affect the rate of reactions involving this product. The multiple reaction mechanism is clearly more complicated than the single toy reaction studied in the present work, and future study should focus on multiple reactions such as the auto-catalytic H$_2$O$_2$ loop described above.  Interestingly, the same H$_2$O$_2$ auto-catalytic loop has also been found responsible of auto-ignition in hydrocarbons \cite{SESHADRI2006131, SAXENA200779}.  

In hydrogen low temperature chemistry, the auto-catalytic loop for H$_2$O$_2$ described above is also controlled by the concentration of HO$_2$, which is controlled by another loop sharing the same potential for non-equilibrium effects, the sequence of reactions 8 and 4.  The high activation energy reaction 8 consumes HO$_2$ and produces H, which is consumed by reaction 4 to produce again the highly energetic HO$_2$.  The net reaction of these two is H$_2$+O$_2$ $\rightarrow$ H$_2$O$_2$, and it proceeds through the sequence of highly energetic HO$_2$ undergoing a highly activated reaction.   This second overall step is also very prone to the same non-equilibrium effects through molecular hotspots.
	
Interestingly, the low temperature auto-ignition of hydrogen is also known experimentally to be very prone to hotspot formation, as first shown by Meyer and Oppenheim \cite{Meyer_1971}.  This is generally attributed to the high sensitivity of the ignition delay to fluctuations, inherited from the  very high activation energies of reactions 6 for the first loop and 8 for the second loop controlling the auto-ignition.  Future study should clearly address whether the propensity for hot spots observed experimentally for this regime is due mainly to coupling to hydrodynamic scale fluctuations \cite{Gardner_2010, MEYER197165, khokhlov2015development, GROGAN20152181}, molecular non-equilibrium effects, or the acceleration of slow reactions by quantum tunneling effects \cite{Eletskii_2005, PhysRevLett.109.183201}, or a combination of these.  

The non-equilibrium high activation and high heat-release effects on auto-ignition are probably most relevant and best studied in detonation waves in both gas and condensed phase, although the shock compression itself may promote other non-equilibrium effects that may couple with the present one.  For example, recent picosecond resolution experiments \cite{armstrong2013} have demonstrated that shocked hydrogen-peroxide reacts on the shock passage time scale of 100 picoseconds, which clearly involves  non-equilibrium effects.	
	
\section{Conclusion}
\label{Conclusion}
Our current results show the existence of two non-equilibrium regimes in a single model reaction for thermal ignition problems.  At high temperatures (low reduced activation energies), the ignition delay was found longer than predicted by standard kinetic theory assuming local equilibrium.  Although the trend was captured by Prigogine and Xhrouet qualitatively, the perturbation model failed at this extreme, as the perturbation assumes slow reactions in the limit of high activation energy. For low temperatures (high activation energies), particularly for high heat release, the ignition delay was found to be shorter than the one predicted by the standard kinetic theory. The perturbation scheme of Prigogine and Mahieu, which incorporates the effect of energy release in the perturbation of the distribution function, accurately captures the effect. 

While the present study clearly highlights the necessity of incorporating non-equilibrium effects in kinetic predictions, particularly for systems where the kinetics are calibrated experimentally in nearly isothermal conditions. It also clearly shows that no existing perturbation scheme provides a uniformly valid approximation. Hence, molecular dynamic results, such as provided in the current paper, can be used to form empirical correction factors to the  existing reaction rates.  

Consideration of the auto-ignition problem in hydrogen-oxygen at low temperatures suggests that non-equilibrium effects involving energy release can be expected in the auto-catalytic loops of H$_2$O$_2$ and HO$_2$ through local sequences of high energy release followed by high activation reaction of these two species.  Future study should focus on these auto-catalytic multiple reactions.  

\section*{Acknowledgments}
\label{Acknowledgments}
M.I.R acknowledges the financial support from the Natural Sciences and Engineering Research Council of Canada (NSERC) through the Discovery Grant "Predictability of detonation wave dynamics in gases: experiment and model development". \\
\bibliographystyle{elsarticle-num}
\bibliography{references} 

\begin{thebibliography}{10}
\expandafter\ifx\csname url\endcsname\relax
  \def\url#1{\texttt{#1}}\fi
\expandafter\ifx\csname urlprefix\endcsname\relax\def\urlprefix{URL }\fi
\expandafter\ifx\csname href\endcsname\relax
  \def\href#1#2{#2} \def\path#1{#1}\fi

\bibitem{Prigogine_1949}
I.~Prigogine, E.~Xhrouet, On the perturbation of {M}axwell distribution
  function by chemical reactions in gases, Physica 15~(11-12) (1949) 913--932.

\bibitem{Prigogine_1950}
I.~Prigogine, M.~Mahieu, Sur la perturbation de la distribution de {M}axwell
  par des reactions chimiques en phase gazeuse, Physica 16~(1) (1950) 51--64.

\bibitem{Williams_1985}
F.~A. Williams, Combustion Theory, The Benjamin/Cummins Publishing Company,
  Menlo Park, California, 1985.

\bibitem{Sanchez_2014}
A.~L. S\'{a}nchez, F.~A. Williams, Recent advances in understanding of
  flammability characteristics of hydrogen, Prog. Energy and Combust. Sci. 41
  (2014) 1--55.

\bibitem{Borisov_1974}
A.~Borisov, On the origin of exothermic centres in gaseous mixtures, Acta
  Astronautica 1~(7) (1974) 909--920.

\bibitem{Gorecki_1987}
J.~Gorecki, J.~Gryko, The adiabatic thermal explosion in a small system:
  Comparison of the stochastic approach with the molecular dynamics simulation,
  J. Stat. Phys. 48~(1-2) (1987) 329--342.

\bibitem{Gorecki_1991}
J.~Gorecki, J.~Popielawski, A.~S. Cukrowski, Molecular-dynamics study on the
  influence of nonequilibrium effects on the rate of chemical reaction,
  Physical Review A 44~(6) (1991) 3791--3795.

\bibitem{Gorecki_2000}
J.~Gorecki, J.~N. Gorecka, Molecular dynamics simulations of nonequilibrium
  rate constant in a model exothermic reaction, Chem. Phys. Lett. 319~(1-2)
  (2000) 173--178.

\bibitem{Lemarchand_2004}
A.~Lemarchand, B.~Nowakowski, Fluctuation-induced and nonequilibrium-induced
  bifurcations in a thermochemical system, Molecular Simulation 30~(11-12)
  (2004) 773--780.

\bibitem{Present_1968}
R.~Present, Chapman-{E}nskog method in chemical kinetics, J. Chem. Phys.
  48~(11) (1968) 4875--4877.

\bibitem{Shizgal_1970}
B.~D. Shizgal, M.~Karplus, Nonequilibrium contributions to the rate of
  reaction. {I}. {P}erturbation of the velocity distribution function, J. Chem.
  Phys. 52~(8) (1970) 4262--4278.

\bibitem{Shizgal_1996}
B.~D. Shizgal, D.~G. Napier, Nonequilibrium effects in reactive systems; the
  effect of reaction products and the validity of the {C}hapman-{E}nskog
  method, Physica A 223 (1996) 50--86.

\bibitem{Chapman_1958}
S.~Chapman, T.~G. Cowling, The mathematical theory of non-uniform gases,
  Cambridge University Press, 1958.

\bibitem{Dean_1985}
A.~M. Dean, Predictions of pressure and temperature effects upon radical
  addition and recombination reactions, American Chemical Society 89 (1985)
  4600--4608.

\bibitem{Westmoreland_1986}
P.~R. Westmoreland, J.~B. Howard, J.~P. Longwell, A.~M. Dean, Combustion and
  pyrolysis reactions by bimolecular {QRRK}, AIChE Journal 32~(12) (1985)
  1971--1979.

\bibitem{Dontgen_2017}
M.~D$\ddot{\text{o}}$ntgen, L.~C. Kr$\ddot{\text{o}}$ger, K.~Leonhard, Hot
  $\beta$-scission of radicals formed via hydrogen abstraction, Proc. Combust.
  Inst. 36 (2017) 135--142.

\bibitem{Dontgen_Feb2017}
M.~D$\ddot{\text{o}}$ntgen, K.~Leonhard, Discussion of the separation of
  chemical and relaxational kinetics of chemically activated intermediates in
  master equation simulations, J. Phys. Chem. 121 (2017) 1563--1570.

\bibitem{Burke_2015}
M.~P. Burke, C.~F. Goldsmith, Y.~Georgievskii, S.~J. Klippenstein, Towards a
  quantitative understanding of the role of non-{B}oltzmann reactant
  distributions in low temperature oxidation, Proc. Combust. Inst. 35 (2015)
  205--213.

\bibitem{Labbe_2017}
N.~J. Labbe, R.~Sivaramakrishnan, C.~F. Goldsmith, Y.~Georgievskii, J.~A.
  Miller, S.~J. Klippenstein, Ramifications of including non-equilibrium
  effects for {HCO} in flame chemistry, Proc. Combust. Inst. 36 (2017)
  525--532.

\bibitem{Goldsmith_2015}
C.~F. Goldsmith, M.~P. Burke, Y.~Georgievskii, S.~J. Klippenstein, Effect of
  non-thermal product energy distributions on ketohydroperoxide decomposition
  kinetics, Proc. Combust. Inst. 35 (2015) 283--290.

\bibitem{Mansour_1992}
M.~M. Mansour, F.~Baras, Microscopic simulation of chemical systems, Physica A
  188~(1) (1992) 253--276.

\bibitem{Sirmas_2017}
N.~Sirmas, M.~I. Radulescu, Thermal ignition revisited with two-dimensional
  molecular dynamics: role of fluctuations in activated collisions, Combustion
  and Flame 13 (2017) 79--88.

\bibitem{Alder_1959}
B.~J. Alder, T.~E. Wainwright, Studies in molecular dynamics. 1. {G}eneral
  method, J. Chem. Phys. 31~(2) (1959) 459--466.

\bibitem{Vincenti_1975}
W.~G. Vincenti, C.~H. Kruger, Introduction to physical gas dynamics, Krieger,
  1975.

\bibitem{doi:10.1080/00102207508946655}
D.~R. Kassoy, A theory of adiabatic, homogeneous explosion from initiation to
  completion, Combustion Science and Technology 10~(1-2) (1975) 27--35.

\bibitem{Pochel_2005}
T.~P$\ddot{\text{o}}$schel, T.~Schwager, Computational granular dynamics:
  models and algorithms, Springer-Verlag, Berlin, Heidelberg, New York, 2005.

\bibitem{Song_1988}
Y.~Song, R.~M. Stratt, E.~A. Mason, The equation of state of hard spheres and
  the approach to random closest packing, J. Chem. Phys 88 (1988) 1126--1133.

\bibitem{Mulero_2008}
A.~Mulero, C.~A. Gal\'{a}n, M.~I. Parra, F.~Cuadros, Equations of state for
  hard spheres and hard disks, Lect. Notes Phys. 753 (2008) 37--109.

\bibitem{Meyer_1971}
J.~W. Meyer, A.~K. Oppenheim, On the shock-induced ignition of explosive gases,
  in: Symposium (International) on Combustion, Vol.~13, Elsevier, 1971, pp.
  1153--1164.

\bibitem{Gardner_2010}
C.~Gardner, Experimental investigation of auto-ignition delay times and soot
  formation in gaseous fuel mixtures, Ph.D. thesis, University of Wales (2010).

\bibitem{SESHADRI2006131}
K.~Seshadri, N.~Peters, G.~Paczko, Rate-ratio asymptotic analysis of
  autoignition of n-heptane in laminar nonpremixed flows, Combustion and Flame
  146~(1) (2006) 131 -- 141.

\bibitem{SAXENA200779}
P.~Saxena, N.~Peters, F.~A. Williams, An analytical approximation for
  high-temperature autoignition times of higher alkanes, Combustion and Flame
  149~(1) (2007) 79 -- 90.

\bibitem{MEYER197165}
J.~W. Meyer, A.~K. Oppenheim, Coherence theory of the strong ignition limit,
  Combustion and Flame 17~(1) (1971) 65 -- 68.

\bibitem{khokhlov2015development}
A.~Khokhlov, Development of hot spots and ignition behind reflected shocks in
  {2H$_2$ + O$_2$}, in: Proceedings of the 25th ICDERS, 2015.

\bibitem{GROGAN20152181}
K.~P. Grogan, M.~Ihme, Weak and strong ignition of hydrogen/oxygen mixtures in
  shock-tube systems, in: Proceedings of the Combustion Institute, Vol.~35,
  2015, pp. 2181 -- 2189.

\bibitem{Eletskii_2005}
A.~V. Eletski, A.~N. Starostin, M.~D. Taran, Quantum corrections to the
  equilibrium rate constants of inelastic processes, Phys. Usp. 48 (2005)
  281--294.

\bibitem{PhysRevLett.109.183201}
A.~V. Drakon, A.~V. Emelianov, A.~V. Eremin, E.~V. Gurentsov, Y.~V.
  Petrushevich, A.~N. Starostin, M.~D. Taran, V.~E. Fortov, Quantum phenomena
  in ignition and detonation at elevated density, Phys. Rev. Lett. 109 (2012)
  183201.

\bibitem{armstrong2013}
M.~R. Armstrong, J.~M. Zaug, N.~Goldman, I.-F.~W. Kuo, J.~C. Crowhurst, W.~M.
  Howard, J.~A. Carter, M.~Kashgarian, J.~M. Chesser, T.~W. Barbee, S.~Bastea,
  Ultrafast shock initiation of exothermic chemistry in hydrogen peroxide, The
  Journal of Physical Chemistry A 117~(49) (2013) 13051--13058.

\end{thebibliography}

\end{document}